\documentclass[twocolumn,showpacs,preprintnumbers,amsmath,amssymb,floatfix]{revtex4}

\usepackage{graphicx}% Include figure files
\usepackage{dcolumn}% Align table columns on decimal point
\usepackage{bm}% bold math

\begin{document}

%%%%
%    Greek Letters
%

\let\a=\alpha      \let\b=\beta       \let\c=\chi        \let\d=\delta
\let\e=\varepsilon \let\f=\varphi     \let\g=\gamma      \let\h=\eta
\let\k=\kappa      \let\l=\lambda     \let\m=\mu
\let\o=\omega      \let\r=\varrho     \let\s=\sigma
\let\t=\tau        \let\th=\vartheta  \let\y=\upsilon    \let\x=\xi
\let\z=\zeta       \let\io=\iota      \let\vp=\varpi     \let\ro=\rho
\let\ph=\phi       \let\ep=\epsilon   \let\te=\theta
\let\n=\nu
\let\D=\Delta   \let\F=\Phi    \let\G=\Gamma  \let\L=\Lambda
\let\O=\Omega   \let\P=\Pi     \let\Ps=\Psi   \let\Si=\Sigma
\let\Th=\Theta  \let\X=\Xi     \let\Y=\Upsilon

%
%%%

%%%
%    Calligraphic letters
%

\def\cA{{\cal A}}                \def\cB{{\cal B}}
\def\cC{{\cal C}}                \def\cD{{\cal D}}
\def\cE{{\cal E}}                \def\cF{{\cal F}}
\def\cG{{\cal G}}                \def\cH{{\cal H}}
\def\cI{{\cal I}}                \def\cJ{{\cal J}}
\def\cK{{\cal K}}                \def\cL{{\cal L}}
\def\cM{{\cal M}}                \def\cN{{\cal N}}
\def\cO{{\cal O}}                \def\cP{{\cal P}}
\def\cQ{{\cal Q}}                \def\cR{{\cal R}}
\def\cS{{\cal S}}                \def\cT{{\cal T}}
\def\cU{{\cal U}}                \def\cV{{\cal V}}
\def\cW{{\cal W}}                \def\cX{{\cal X}}
\def\cY{{\cal Y}}                \def\cZ{{\cal Z}}
%
%%%%

\newcommand{\Ns}{N\hspace{-4.7mm}\not\hspace{2.7mm}}
\newcommand{\qs}{q\hspace{-3.7mm}\not\hspace{3.4mm}}
\newcommand{\ps}{p\hspace{-3.3mm}\not\hspace{1.2mm}}
\newcommand{\ks}{k\hspace{-3.3mm}\not\hspace{1.2mm}}
\newcommand{\des}{\partial\hspace{-4.mm}\not\hspace{2.5mm}}
\newcommand{\desco}{D\hspace{-4mm}\not\hspace{2mm}}

%%%%

%\draft command makes pacs numbers print
%\draft

\title{ Enhanced $K_L \to \pi^0\nu\bar\nu$ from Direct $CP$ Violation in
 $B \to K\pi$  with Four Generations
}
\author{Wei-Shu Hou$^{a,b}$}
%\email{wshou@phys.ntu.edu.tw}
\author{Makiko Nagashima$^a$}
%\email{}
\author{Andrea Soddu$^{a,c}$}
%\email{asoddu@hep1.phys.ntu.edu.tw}
\affiliation{
$^a$Department of Physics, National Taiwan
 University, Taipei, Taiwan 106, R.O.C. \\
$^b$Stanford Linear Accelerator Center,
 Stanford, California 94309, U.S.A. \\
$^c$Department of Particle Physics, Weizmann Institute
 of Science, Rehovot 76100, Israel
}
\date{\today}

\begin{abstract}
Recent $CP$ violation results in $B$ decays suggest that $Z$
penguins may have large weak phase. This can be realized by the
four generation (standard) model.
% with sizable and near imaginary $V_{t^\prime s}^* V_{t^\prime b}$.
Concurrently, $B\to X_s\ell^+\ell^-$ and $B_s$ mixing allow for
sizable $V_{t^\prime s}^* V_{t^\prime b}$ only if it is nearly
imaginary. Such large effects in $b \leftrightarrow s$ transitions
would affect $s\leftrightarrow d$ transitions, as kaon constraints
would demand $V_{t^\prime d} \neq 0$. Using $\Gamma(Z\to b\bar b)$
to bound $\vert V_{t^\prime b}\vert$, we infer sizable $\vert
V_{t^\prime s}\vert \lesssim \vert V_{t^\prime b}\vert \lesssim
\vert V_{us}\vert$. Imposing $\varepsilon_K$, $K^+ \to
\pi^+\nu\bar\nu$ and $\varepsilon^\prime/\varepsilon$ constraints,
we find $V_{t^\prime d}^* V_{t^\prime s} \sim$ few $\times
10^{-4}$ with large phase, enhancing $K_L \to \pi^0\nu\bar\nu$ to
$5\times 10^{-10}$ or even higher.
%A larger hadronic parameter$R_6$ is needed for
%$\epsilon^\prime/\epsilon$ to be satisfied.
Interestingly, $\Delta m_{B_d}$ and $\sin2\Phi_{B_d}$ are not much
affected, as $\vert V_{t^\prime d}^* V_{t^\prime b}\vert \ll \vert
V_{td}^* V_{tb}\vert \sim 0.01$.
\end{abstract}

\maketitle
%\narrowtext

%\section{Introduction}

Just 3 years after $CP$ violation (CPV) in the B system was
established, direct $CP$ violation (DCPV) was also observed in
$B^0\to K^+\pi^-$ decay, ${\cal A}_{K^+\pi^-} \sim -0.12$. A
puzzle emerged, however, that the charged $B^+\to K^+\pi^0$ mode
gave no indication of DCPV, and is in fact a little positive,
${\cal A}_{K^+\pi^0} \gtrsim 0$. Currently, ${\cal A}_{K^+\pi^0} -
{\cal A}_{K^+\pi^-} \simeq 0.16$, and differs from zero with
3.8$\sigma$ significance \cite{AKpiAKpi0}.

The amplitude ${\cal M}_{K^+\pi^-} \simeq P + T$ is dominated by
the strong penguin ($P$) and tree ($T$) contributions, while the
main difference $\sqrt2{\cal M}_{K^+\pi^0} - {\cal M}_{K^+\pi^-}
\simeq P_{\rm EW} + C$ is from electroweak penguin (EWP, or
$P_{\rm EW}$) and color-suppressed tree ($C$) contributions which
are subdominant. Thus, ${\cal A}_{K^+\pi^0} \sim {\cal
A}_{K^+\pi^-}$ was anticipated by all models.
As data indicated otherwise, it has been stressed~\cite{LargeC}
that the $C$ term could be much larger than previously thought,
effectively cancelling against the CPV phase in $T$, leading to
${\cal A}_{K\pi^0} \to 0$. While this may well be realized, a very
large $C$ (especially if ${\cal A}_{K\pi^0}
> 0$) would be a surprise in itself.

In a previous paper~\cite{Kpi0HNS}, we explored the possibility of
New Physics (NP) effects in $P_{\rm EW}$, in particular in the 4
generation standard model (SM4, with SM3 for 3 generations). A
sequential $t^\prime$ quark could affect $P_{\rm EW}$ most
naturally for two reasons. On one hand, the associated
Cabibbo-Kobayashi-Maskawa (CKM) matrix element product
$V_{t^\prime s}^*V_{t^\prime b}$ %\equiv r_{sb}e^{i\phi_{sb}}$
could be large and imaginary; on the other hand, it is well known
that $P_{\rm EW}$ is sensitive to $m^2_{t^\prime}$ in amplitude,
and heavy $t^\prime$ does not decouple.

Using the PQCD factorization approach at leading
order~\cite{PQCDKpiLO}, which successfully predicted ${\cal
A}_{K^+\pi^-} < -0.1$ (and $C$ was not inordinately large), we
showed that ${\cal A}_{K^+\pi^0} \gtrsim 0$ called for sizable
$m_{t^\prime} \gtrsim 300$ GeV and large, nearly imaginary
$V_{t^\prime s}^*V_{t^\prime b}$.
As the $m_{t^\prime}$ dependence is similar, we also showed that
data on $B \to X_s\ell^+\ell^-$ and $B_s$ mixing concurred, in the
sense that large $t^\prime$ effect is allowed {\it only if}
$V_{t^\prime s}^*V_{t^\prime b}$ is nearly imaginary. Applying the
latter two constraints, however, $m_{t^\prime}$ and $V_{t^\prime
s}^*V_{t^\prime b}$ become highly constrained. In the following,
we will take~\cite{Kpi0HNS}
\begin{equation}
m_{t^\prime} \cong 300\ {\rm GeV},\ V_{t^\prime s}^*V_{t^\prime b}
\equiv r_{sb}\,e^{i\phi_{sb}} \simeq 0.025\, e^{i\,70^\circ},
\label{eq:rsbphisb}
\end{equation}
as exemplary %, if not somewhat extreme,
values for realizing
${\cal A}_{K^+\pi^0} - {\cal A}_{K^+\pi^-} \gtrsim 0.10$, without
recourse to a large $C$ contribution.

Comparing with $\vert V_{cs}V_{cb}\vert \simeq 0.04$, $r_{sb} \sim
0.025$ is quite sizable.
%A large effect is needed, however, since
%one is moving ${\cal A}_{K^+\pi^0}$ from $\sim -0.10$ to zero with
%loop contributions.
%
In our $b\to s$ study, we had assumed~\cite{Kpi0HNS} $V_{t^\prime
d} \to 0$ out of convenience, so as to decouple from $b\to d$ and
$s\to d$ concerns. The main purpose of this note, however, is to
show that, in view of the large $r_{sb}$ and $\phi_{sb}$ values
given in Eq.~(\ref{eq:rsbphisb}), $V_{t^\prime d} = 0$ is
untenable, and one must explore $s\to d$ and $b\to d$
implications. The reasoning is as follows. Since a rather large
impact on $V_{ts}^*V_{tb}$ is implied by Eq.~(\ref{eq:rsbphisb}),
if one sets $V_{t^\prime d} = 0$, then $V^*_{td}V_{ts}$ would
still be rather different from SM3 case. With our current
knowledge of $m_t$, the $\varepsilon_K$ parameter would deviate
from the well measured experimental value. Thus, a finite
$V_{t^\prime d}$ is needed to tune for $\varepsilon_K$.

We find that the kaon constraints that are sensitive to $t^\prime$
(i.e. $P_{\rm EW}$-like), viz. $K^+\to \pi^+\nu\bar\nu$, $K_L\to
\mu^+\mu^-$, $\varepsilon_K$, and $\varepsilon^\prime/\varepsilon$
can all be satisfied.
%, provided that the hadronic matrix element
%$R_6$ for $\varepsilon^\prime/\varepsilon$ is different from the
%value that is typically~\cite{BurasNLO} used for SM3.
Interestingly, once kaon constraints are satisfied, we find little
impact is implied for $b\leftrightarrow d$ transitions, such as
$\Delta m_{B_d}$ and $\sin2\Phi_{B_d}$. That is, $V_{t^\prime d}
\to 0$ works approximately for $b\to d$ transitions, for current
level of experimental sensitivity. The main outcome for $s\to d$
and $b\to d$ transitions is the enhancement of $K_L\to
\pi^0\nu\bar\nu$ mode by an order of magnitude or more, to beyond
$5\times 10^{-10}$.

%%%%
\begin{figure}[b!]
\smallskip  %\smallskip
\includegraphics[width=1.5in,height=0.7in,angle=0]{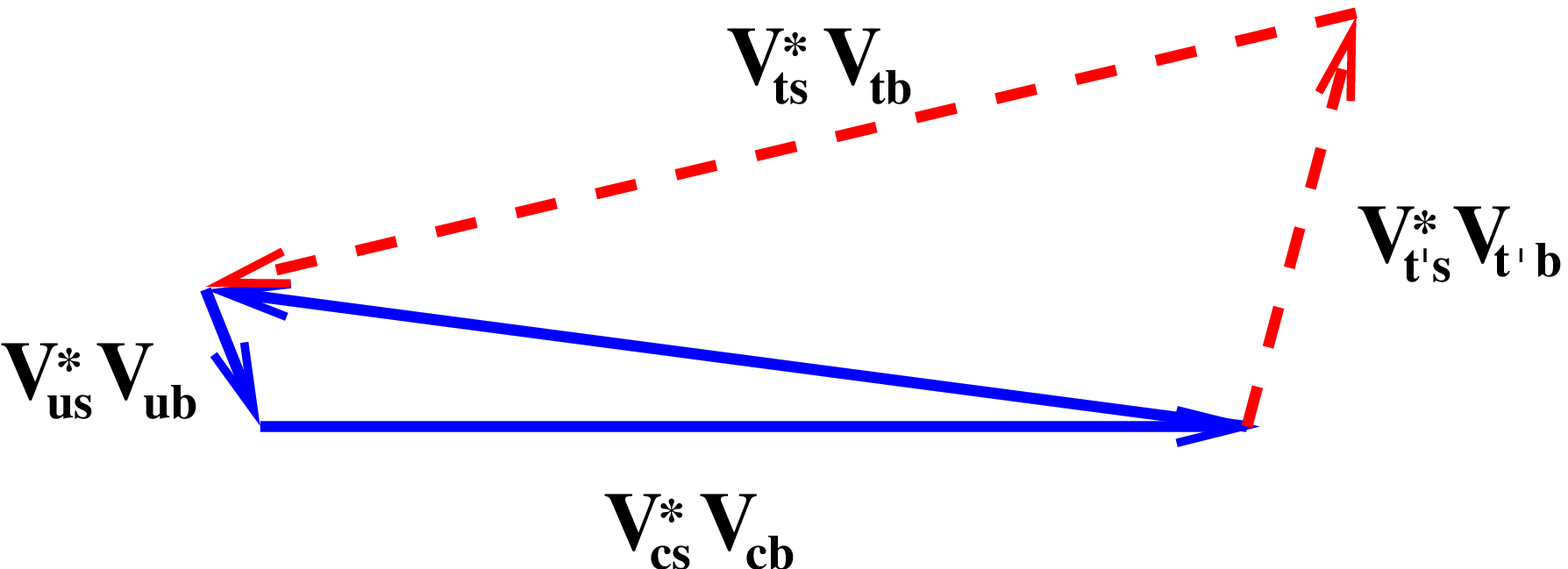}
\hskip0.1cm
\includegraphics[width=1.5in,height=0.7in,angle=0]{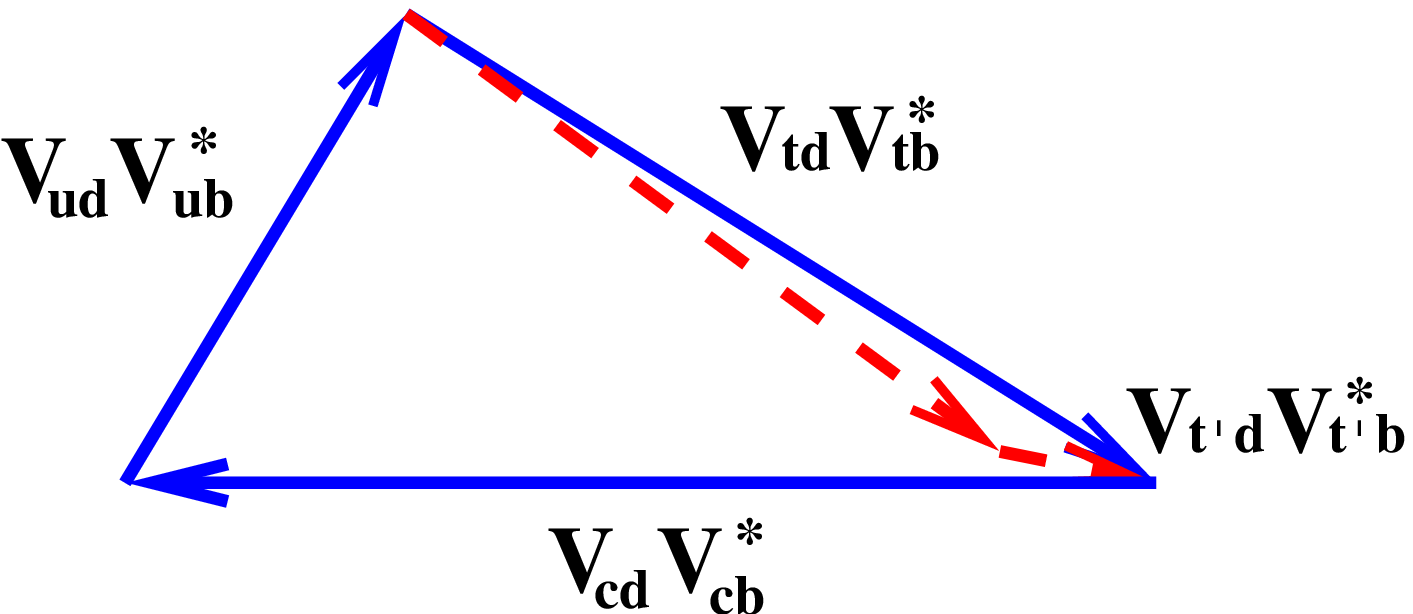}
 \vskip-0.2cm
%\smallskip\smallskip%\smallskip\smallskip\smallskip
 \caption{
 Unitarity quadrangles of
  (a) Eq.~(\ref{eq:sbquad}), with $\vert V_{us}^*V_{ub}\vert$ exaggerated;
  (b) Eq.~(\ref{eq:dbquad}), where actual scale is $\sim 1/4$ of (a).
    Adding $V_{t^\prime s}^*V_{t^\prime b}$ (dashed) according to
    Eq.~(\ref{eq:rsbphisb}) drastically changes the invariant phase and
    $V_{ts}^*V_{tb}$ from the SM3 triangle (solid),
    but from Eq.~(\ref{eq:db_vals}), the dashed lines for $V_{td}V^*_{tb}$
    and $V_{t^\prime d}V^*_{t^\prime b}$ can hardly be distinguished from
    SM3 case.
 \label{fig:Quadr}
}
\end{figure}
%%%%

%\section{$4\times 4$ Unitarity and Parametrization}

With four generations, adding $V_{t^\prime s}^*V_{t^\prime b}$
extends the familiar unitarity triangle relation into a
quadrangle,
\begin{equation}
V_{us}^*V_{ub} + V_{cs}^*V_{cb} + V_{ts}^*V_{tb} + V_{t^\prime
s}^*V_{t^\prime b} = 0.
\label{eq:sbquad}
\end{equation}
Using SM3 values for $V_{us}^*V_{ub}$, $V_{cs}^*V_{cb}$ (validated
later by our $b\to d$ study), since they are probed in multiple
ways already, and taking $V_{t^\prime s}^*V_{t^\prime b}$ as given
in Eq.~(\ref{eq:rsbphisb}), we depict Eq.~(\ref{eq:sbquad}) in
Fig.~1(a). The solid, rather squashed triangle is the usual
$V_{us}^*V_{ub} + V_{cs}^*V_{cb} + V_{ts}^*V_{tb} = 0$ in SM3.
Given the size and phase of $V_{t^\prime s}^*V_{t^\prime b}$, one
sees that the invariant phase represented by the area of the
quadrangle is rather large, and $V_{ts}^*V_{tb}$ picks up a large
imaginary part, which is very different from SM3 case. Such large
effect in $b\to s$ would likely spill over into $s\to d$
transitions, since taking $V_{tb}$ as real and of order 1, one
immediately finds the strength and complexity of $V_{td}^*V_{ts}$
would be rather different from SM3, and one would need
$V_{t^\prime d}^*V_{t^\prime s} \neq 0$ to compensate for the well
measured value for $\varepsilon_K$.

Note from Fig. 1(a) that the usual approximation of dropping
$V_{us}^*V_{ub}$ in the loop remains a good one. To face $s\to d$
and $b\to d$ transitions, however, one should respect unitarity of
the $4\times 4$ CKM matrix $V_{\rm CKM}$. We adopt the
parametrization in Ref.~\cite{HSS87} where the third column and
fourth row is kept simple. This is suitable for $B$ physics, as
well as for loop effects in kaon sector. With $V_{cb}$, $V_{tb}$
and $V_{t^\prime b}$ defined as real, one keeps the SM3 phase
convention for $V_{ub}$, now defined as
\begin{equation}
\arg V_{ub}^* = \phi_{ub},
 \label{eq:phiub}
\end{equation}
which is usually called $\phi_3$ or $\gamma$ in SM3.
We take $\phi_{ub} = 60^\circ$ as our nominal
value~\cite{varphi3}.
This can in principle be measured through tree level processes
such as the $B\to DK$ Dalitz method~\cite{DKDalitz}. The two
additional phases are associated with $V_{t^\prime s}$ and
$V_{t^\prime d}$, and for the rotation angles we follow the PDG
notation~\cite{PDG}.
To wit, we have
\begin{eqnarray}
V_{t^\prime d} &=& -c_{24}c_{34}s_{14} e^{-i\phi_{db}},
\label{eq:Vt'd} \\
V_{t^\prime s} &=& -c_{34}s_{24} e^{-i\phi_{sb}},\
\label{eq:Vt's} \\
V_{t^\prime b} &=& -s_{34},
\label{eq:Vt'b}
\end{eqnarray}
while $V_{t^\prime b^\prime} = c_{14}c_{24}c_{34}$,
$V_{tb} = c_{13}c_{23}c_{34}$,
$V_{cb} = c_{13}c_{34}s_{23}$ are all real. With this convention
for rotation angles, from Eq.~(\ref{eq:phiub}) we have
$V_{ub} = c_{34}s_{13}e^{-i\phi_{ub}}$.

Analogous to Eq.~(\ref{eq:rsbphisb}), we also make the heuristic
but redundant definition of
\begin{equation}
V_{t^\prime d}^*V_{t^\prime b} \equiv r_{db}\,e^{i\,\phi_{db}},\
V_{t^\prime d}^*V_{t^\prime s} \equiv r_{ds}\,e^{i\,\phi_{ds}},
\label{eq:dbsb}
\end{equation}
as these combinations enter $b\to d$ and $s\to d$ transitions.
Inspection of Eqs.~(\ref{eq:rsbphisb}), (\ref{eq:Vt'd}
%), (\ref{eq:Vt's}), and (
--\ref{eq:Vt'b}) gives the relations
\begin{equation}
r_{db}r_{sb} = r_{ds}s_{34}^2,\
\phi_{ds} = \phi_{db} - \phi_{sb}.
\label{eq:rdsphids}
\end{equation}
As we shall see, $s\to d$ transitions are much more stringent than
$b\to d$ transitions, hence we shall turn to constraining $r_{ds}$
and $\phi_{ds}$.

%\section{ $Z\to b\bar b$ Constraint
% Combining $b \leftrightarrow s$:
% $\,V_{t^\prime s} \sim V_{t^\prime b} \sim \lambda$
%}

Before turning to the kaon sector, we need to infer what value to
use for $s_{34} = \vert V_{t^\prime b}\vert$, as this can still
affect the relevant physics through unitarity. Fortunately, we
have some constraint on $s_{34}$ from $Z\to b\bar b$ width, which
receives special $t$ (and hence $t^\prime$) contribution compared
to other $Z\to q\bar q$, and is now suitably well measured.

Following Ref.~\cite{Yanir} and using
% $\bar{m}_t(m_t) = 166$ GeV,
$m_{t^\prime} = 300$ GeV, we find
\begin{equation}
\vert V_{tb}\vert^2 + 3.4 \vert V_{t^\prime b}\vert^2 < 1.14.
\label{eq:Zbb}
\end{equation}
Since all $c_{ij}$s except perhaps $c_{34}$ would still likely be
close to 1, we infer that $s_{34} \lesssim 0.25$.
We take the liberty to nearly saturate this bound ($\Gamma(Z\to
b\bar b)$ is close to 1$\sigma$ above SM3 expectation), by {\it
imposing}
\begin{equation}
s_{34} \simeq 0.22,
\label{eq:s34}
\end{equation}
to be close to the Cabibbo angle, $\lambda \equiv \vert V_{us}
\vert \cong 0.22$.
Note that Eq.~(\ref{eq:s34}) is somewhat below the expectation of
``maximal mixing" of $s_{34}^2 \sim 1/2$ between third and fourth
generations. Combining it with Eq.~(\ref{eq:rsbphisb}), one gets
$\vert V_{t^\prime s} \vert \sim 0.11 \sim \lambda/2$. Its
strength would grow if a lower value of $s_{34} \lesssim \lambda$
is chosen, which would make even greater impact on $s\to d$
transitions.

%\section{ $K^+\to \pi^+\nu\bar\nu,\ K_L\to \mu^+\mu^-,\
%        \varepsilon_K$ and
%        $\varepsilon^\prime/\varepsilon$ Constraints
%         %$\Rightarrow$ $r_{ds}$, $\phi_{ds}$
%         }

Using current values~\cite{PDG} of $V_{cb}$ and $V_{ub}$ as input
and respecting full unitarity, we now turn to the kaon constraints
of $K^+\to \pi^+\nu\bar\nu$, $\varepsilon_K$, $K_L\to \mu^+\mu^-$,
and $\varepsilon^\prime/\varepsilon$. The first two are
short-distance (SD) dominated, while the last two suffer from
long-distance (LD) effects.

Let us start with $K^+\to \pi^+\nu\bar\nu$. The first observed
event~\cite{E787} by E787 suggested a sizable rate hence hinted at
NP. The fourth generation would be a good candidate, since the
process is dominated by the $Z$ penguin. Continued running,
including E949 data (unfortunately not greatly improving
accumulated luminosity), has yielded overall 3 events, and the
rate is now ${\cal B}(K^+\to \pi^+\nu\bar\nu) =
(1.47^{+1.30}_{-0.89})\times 10^{-10}$~\cite{E949}. This is still
somewhat higher than the SM3 expectation of order $0.8\times
10^{-10}$.

Defining $\lambda_{q}^{ds} \equiv V_{qd}V_{qs}^{*}$ and using the
formula~\cite{BSU}
\begin{eqnarray}
{\cal B}(K^+ \rightarrow \pi^+ \nu \bar{\nu}) &=& \kappa_+
\left|\frac{\lambda_c^{ds}}{|V_{us}|}P_c +
\frac{\lambda_t^{ds}}{|V_{us}|^5} \eta_t X_0(x_t) \right.
\nonumber
\\
&& \left . + \frac{\lambda_{t^{\prime}}^{ds}}{|V_{us}|^5}
\eta_{t^{\prime}} X_0(x_{t^{\prime}}) \right|^2,
\label{eq:pinunu}
\end{eqnarray}
we plot in Fig.~2 the allowed range (valley shaped shaded region)
of $r_{ds}$--$\phi_{ds}$ for the 90\% confidence level (C.L.)
bound of ${\cal B}(K^+ \rightarrow \pi^+ \nu \bar{\nu}) <
3.6\times 10^{-10}$.
We have used~\cite{BSU} $\kappa_+=(4.84 \pm 0.06)\times
10^{-11}\times(0.224/|V_{us}|)^8$ and $P_c=(0.39 \pm
0.07)\times(0.224/|V_{us}|)^4$. We take the QCD correction factors
$\eta_{t^{(\prime)}} \sim 1$, and $X_0(x_{t^{(\prime)}})$
evaluated for $m_t = 166$ GeV and $m_{t^\prime} = 300$ GeV.
We see that $r_{ds}$ up to $7 \times 10^{-4}$ is possible, which
is not smaller than the SM3 value of $4 \times 10^{-4}$ for $\vert
V_{td}^*V_{ts}\vert$.
% This is in part due to the current central value for
%${\cal B}(K^+\to \pi^+\nu\bar\nu)$ being higher than predicted by SM3.

The SD contribution to $K_L \to\mu^+\mu^-$ is also of interest.
The $K_L \to\mu^+\mu^-$ rate is saturated by the absorptive
$K_L\to \gamma\gamma \to \mu^+\mu^-$, while the off-shell photon
contribution makes the SD contribution hard to constrain. To be
conservative, we use the experimental bound of ${\cal B}(K_L
\to\mu^+\mu^-)_{\rm SD} < 3.7 \times 10^{-9}$~\cite{E871}. It is
then in general less stringent than $K^+\to \pi^+\nu\bar\nu$,
although the generic constraint on $r_{ds}$ drops slightly. We do
not plot this constraint in Fig.~2.

\begin{figure}[t!]
%\smallskip  %\smallskip
\includegraphics[width=2.2in,height=1.6in,angle=0]{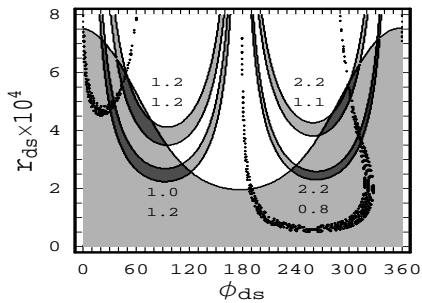}
%\includegraphics[width=1.61in,height=1.61in,angle=0]{epsprOVERepsR6R8}
%\smallskip \hspace{5.mm} \vspace{20.mm}
 \caption{
 Allowed region from $K^+ \rightarrow \pi^+ \nu \bar{\nu}$ (valley
 shaped shaded region), $\varepsilon_K$ (simulated dots) and
 $\varepsilon^\prime/\varepsilon$ (elliptic rings)
  in $r_{ds}$ and $\phi_{ds}$ plane, as described in text, where
  $V^*_{t^\prime d}V_{t^\prime s} \equiv r_{ds}\, e^{i\phi_{ds}}$.
 For $\varepsilon^\prime/\varepsilon$,
  the rings on upper right correspond to
   $R_6 = 2.2$, and $R_8 = 0.8$, $1.1$ (bottom to top),
  and on upper left, $R_6 = 1.0$, $1.2$ (bottom to top), $R_8 =1.2$.
% (b) Allowed $R_6$-$R_8$ range consistent with other constraints.
%  The upper two lines are for $\phi_{ds} = +60^\circ$ and $r_{ds} =
%  3$ (upper) and $5\times 10^{-4}$ (lower), and
%  the lower two lines are for $\phi_{ds} = -60^\circ$ and $r_{ds} =
%  3$ (lower) and $5\times 10^{-4}$ (upper), respectively.
 }
 \label{fig:Fig2}
\end{figure}

The rather precisely measured CPV parameter $\varepsilon_K =
(2.284 \pm 0.014) \times 10^{-3}$~\cite{PDG} is predominantly SD.
It maps out rather thin slices of allowed regions on the
$r_{ds}$--$\phi_{ds}$ plane, as illustrated by dots in Fig.~2,
where we use the formula of Ref.~\cite{Yanir} and follow the
treatment. Note that $r_{ds}$ up to $7\times 10^{-4}$ is still
possible, for several range of values for $\phi_{ds}$. This is the
aforementioned effect that extra CPV effects due to large
$\phi_{sb}$ and $r_{sb}$ now have to be tuned by $t^\prime$ effect
to reach the correct $\varepsilon_K$ value.
We have checked that $\Delta m_K$ makes no additional new
constraint.

The DCPV parameter, Re$\,(\varepsilon^\prime/\varepsilon)$, was
first measured in 1999~\cite{eprime}, with current value at $(1.67
\pm 0.26) \times 10^{-3}$~\cite{PDG}.
%, where we follow PDG and use a scale factor for the error.
It depends on a myriad of hadronic parameters, such as $m_s$,
$\Omega_{\rm IB}$ (isospin breaking), and especially the
non-perturbative parameters $R_6$ and $R_8$, which are related to
the hadronic matrix elements of the dominant strong and
electroweak penguin operators. With associated large
uncertainties, we expect $\,\varepsilon^\prime/\varepsilon$ to be
rather accommodating, but for specific values of $R_6$ and $R_8$,
some range for $r_{ds}$ and $\phi_{ds}$ is determined.
%It is intriguing to find that this is not the case.

We use the formula
\begin{equation}
{\rm Re}\,\frac{\varepsilon^{\prime}}{\varepsilon} = {\rm
Im}\,(\lambda_c^{ds}) P_0 + {\rm Im}\,(\lambda_t^{ds}) F(x_t) +
{\rm Im}\,(\lambda_{t^{\prime}}^{ds}) F(x_{t^{\prime}}),
 \label{eq:epsilonprime}
\end{equation}
where $F(x)$ is given by
\begin{equation}
F(x) = P_{X}X_0(x) + P_{Y}Y_0(x) + P_Z Z_0(x) +P_E E_0(x).
% \nonumber
\end{equation}
The SD functions $X_0$, $Y_0$, $Z_0$ and $E_0$ can be found, for
example, in Ref.~\cite{BurasNLO}, and the coefficients $P_i$ are
given in terms of $R_6$ and $R_8$ as
\begin{equation}
P_i = r_i^{(0)} + r_i^{(6)} R_6 + r_i^{(8)} R_8,
\end{equation}
which depends on LD physics. We differ from Ref.~\cite{BurasNLO}
by placing $P_0$, multiplied by ${\rm Im}\,(\lambda_c^{ds})$,
explicitly in Eq.~(\ref{eq:epsilonprime}). In SM4, one no longer
has the relation ${\rm Im}{\lambda_c^{ds}}=-{\rm
Im}{\lambda_t^{ds}}$ that makes
Re$\,(\varepsilon^\prime/\varepsilon)$ proportional to ${\rm
Im}(\lambda_t^{ds})$. We take the $r_i^{(j)}$ values from
Ref.~\cite{BurasNLO} for $\Lambda_{\overline{MS}}^{(4)} = 310\
{\rm MeV}$,
%
%In Table 1 we give the coefficients
%$r_i^{(0)},r_i^{(6)}$ and $r_i^{(8)}$ for
%$\Lambda_{\overline{MS}}^{(4)}=310 {\rm MeV}$.
%\begin{table}
%\caption{The coefficients $r_i^{(0)},r_i^{(6)}$ and $r_i^{(8)}$
%for $\Lambda_{\overline{MS}}^{(4)}=310 {\rm MeV}$ in NDR scheme. }
%\begin{center}
%\setlength{\tabcolsep}{12pt}
%\begin{ruledtabular}
%\begin{tabular}{cccc}
%$i$ &$r_i^{(0)}$&$r_i^{(6)}$&$r_i^{(8)}$
%\\ \hline \\
%$0$ & $ 3.574$  & $-16.552$  & $-1.805$   \\
%$X_0$ & $0.574$  & $0.030$  & $0$    \\
%$Y_0$ & $0.403$  & $0.119$  & $0$    \\
%$Z_0$ & $0.714$  & $-0.023$  & $-12.510$   \\
%$E_0$ & $0.213$  & $-1.909$  & $0.550$   \\
%\end{tabular}
%\end{ruledtabular}
%\end{center}
%\end{table}
%
 %in NDR scheme.
but reverse the sign of $r_0^{(j)}$ for above mentioned reason.
Note that Re$\,(\varepsilon^\prime/\varepsilon)$ depends linearly
on $R_6$ and $R_8$. For fixed SD parameters $m_{t^\prime}$ and
$\lambda^{ds}_{t^\prime} = V_{t^\prime d} V^*_{t^\prime s}$, one
may adjust for solutions to $K^+\to \pi^+\nu\bar\nu$ and
$\varepsilon_K$.

For the ``standard"~\cite{BurasNLO} parameter range of $R_6 = 1.23
\pm 0.16$ and $R_8 = 1.0 \pm 0.2$, we find $R_8 \sim 1.2$ and $R_6
\sim 1.0$--$1.2$ allows for solutions at $r_{ds} \sim (5$--$6)
\times 10^{-4}$ with $\phi_{ds} \sim +(35^\circ$--$50^\circ)$, as
illustrated by the elliptic rings on upper left part of Fig.~2.
For $R_6 = 2.2 \pm 0.4$ found~\cite{Bijnens} in $1/N_C$ expansion
at next-to-leading order (and chiral perturbation theory at
leading order), within SM3 one has trouble giving the correct
Re$\,(\varepsilon^\prime/\varepsilon)$ value. However, for SM4,
solutions exist for $R_6 \sim 2.2$ and $R_8 = 0.8$--$1.1$, for
$r_{ds} \sim (3.5$--$5) \times 10^{-4}$ and $\phi_{ds} \sim
-(45^\circ$--$60)^\circ$, as illustrated by the elliptic rings on
upper right part of Fig.~2.
%
%For $R_6 = 2.1 \pm 1.1$ and $R_8 = 2.2 \pm 0.4$~\cite{HPdR}, where
%dominant light quark corrections in the large $N_C$ limit are
%considered, we find $R_6 \sim 2.1$ and $R_8 \sim 2.2$ also gives
%$r_{ds} \sim 5 \times 10^{-4}$ and $\phi_{ds} \sim -60^\circ$ as
%solutions.
%
We will take
\begin{equation}
r_{ds} \sim 5\times 10^{-4},\ \ \phi_{ds} \sim -60^\circ\  {\rm
or}\ +35^\circ,
\label{eq:ds_vals}
\end{equation}
as our two nominal cases that satisfy all kaon constraints. The
corresponding values for $R_6$ and $R_8$ can be roughly read off
from Fig.~2.
%To illustrate this, we show in Fig.~2(b) the
%correlated allowed range in $R_6$--$R_8$ that satisfy all
%constraints, for $r_{ds} \sim 3$ and $5\times 10^{-4}$, and
%$\phi_{ds} = \pm 60^\circ$. Thus, the range of solution space from
%kaon sector is roughly as in Eq.~(\ref{eq:ds_vals}), even though
%there is much uncertainty in $R_6$ and $R_8$.
%
We stress again that these values should be taken as exemplary.

\begin{figure}[b!]
\smallskip  %\smallskip
\hspace{-1mm}\vspace{-2.mm}
\includegraphics[width=1.59in,height=1.02in,angle=0]{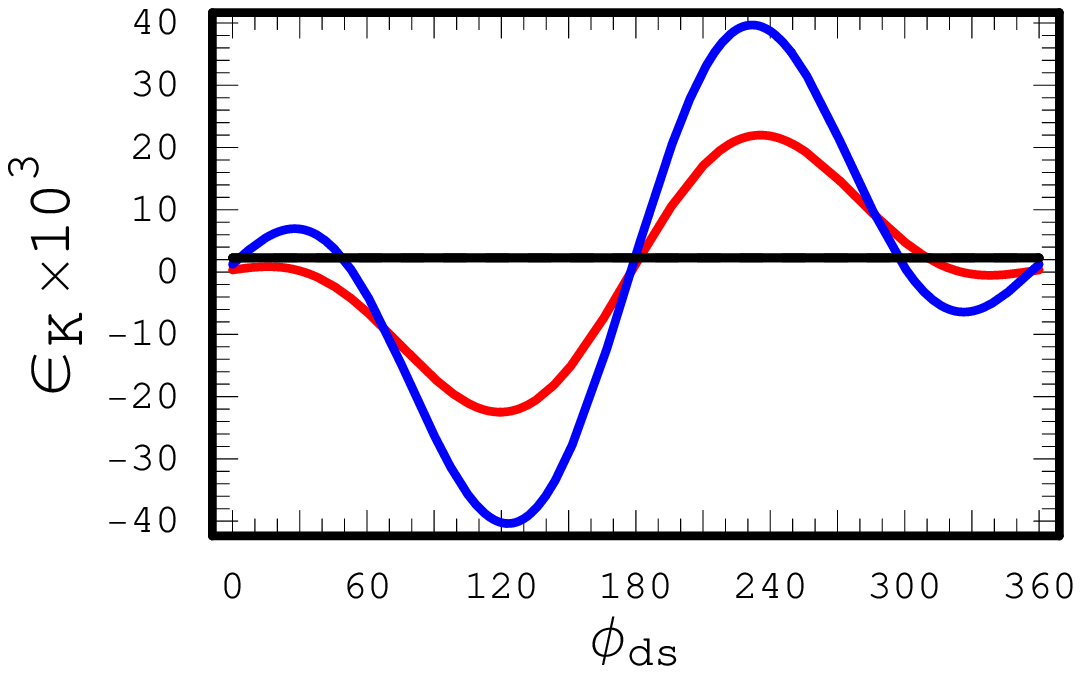}
\hspace{-1.1mm}\vspace{-2.mm}
\includegraphics[width=1.6in,height=1.0in,angle=0]{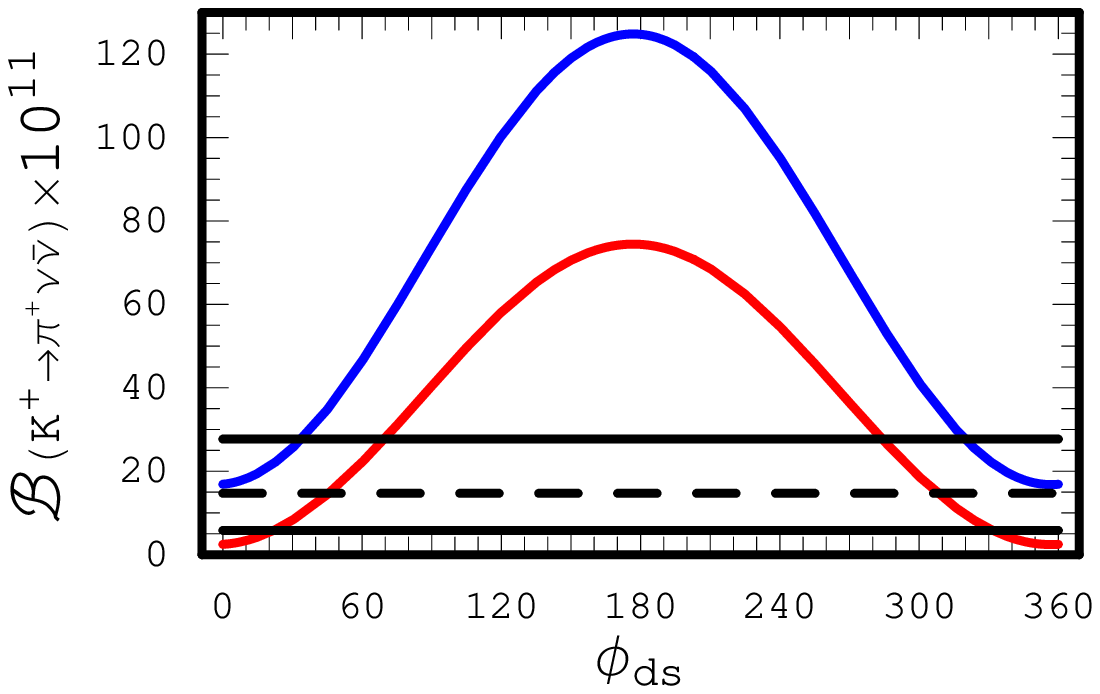}
\vspace{5mm} \hspace{2.0mm}
\includegraphics[width=1.62in,height=1.0in,angle=0]{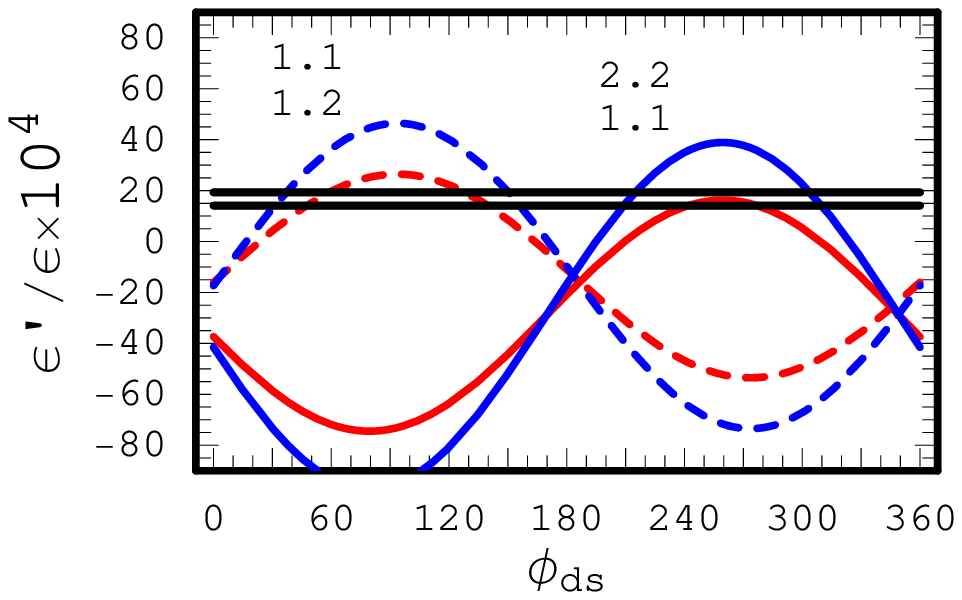}
\vspace{7mm}\hspace{-1mm}
\includegraphics[width=1.6in,height=1.0in,angle=0]{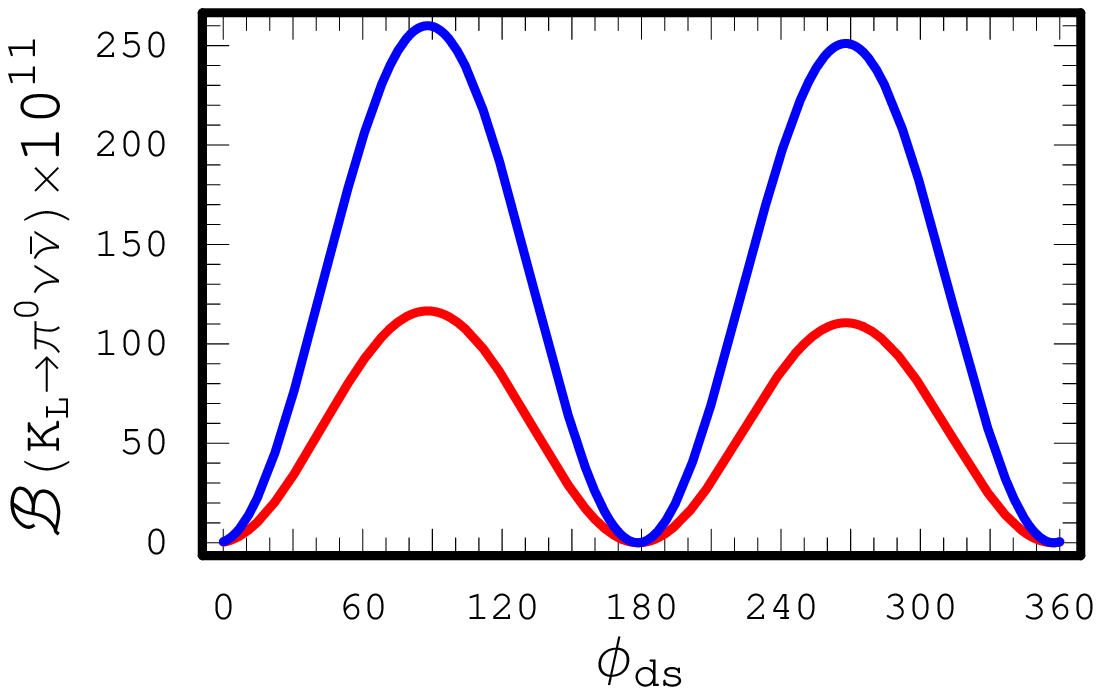}
\vspace{-1mm}
 %\smallskip\smallskip%\smallskip\smallskip\smallskip
 \vskip-0.1cm
\caption{
  (a) $\varepsilon_K$,
  (b) ${\cal B}(K^+\to \pi^+\nu\bar\nu)$,
  (c) Re$\,(\varepsilon^\prime/\varepsilon)$ and
  (d)~${\cal B}(K_L \to \pi^0\nu\bar\nu)$
   vs $\phi_{ds}$, for $r_{ds} = 4$ and $6\times 10^{-4}$
   and $m_{t^\prime} = 300$ GeV. Larger $r_{ds}$ gives stronger
   variation, and horizontal bands are current ($1\sigma$) experimental
   range~\cite{PDG} (the bound for (d) is outside the plot).
  For (c), solid (dashed) lines are for $R_6 = 2.2$, $R_8 = 1.1$
   ($R_6 = 1.1$, $R_8 = 1.2$).
  }
 \label{fig:Fig3}
%\vspace{-2mm}
\end{figure}

To illustrate in a different way, we plot $\varepsilon_K$, ${\cal
B}(K^+\to \pi^+\nu\bar\nu)$ and
Re$\,(\varepsilon^\prime/\varepsilon)$ vs $\phi_{ds}$ in Figs.
3(a), (b) and (c), respectively, for $r_{ds}= 4$ and $6 \times
10^{-4}$. The current $1\sigma$ experimental range is also
illustrated. In Fig.~3(c), we have illustrated with $R_6 = 1.1$,
$R_8 = 1.2$~\cite{BurasNLO} and $R_6 = 2.2$, $R_8 =
1.1$~\cite{Bijnens}. For the former (latter) case, the variation
is enhanced as $R_6$ ($R_8$) drops.
%and $R_6 = 2.2$, $R_8 = 1.2$
%$0.9$ and $1.0$, and we can see the
%This in turn implies ${\cal B}(K^+\to \pi^+\nu\bar\nu)$ should be
%at the higher side of experimental range, $\gtrsim 1.2\times
%10^{-10}$.

%\section{ Enhanced $K_L \to \pi^0\nu\bar\nu$}

It is interesting to see what are the implications for the CPV
decay $K_L \to \pi^0\nu\bar\nu$. The formula for ${\cal B}(K_L \to
\pi^0\nu\bar\nu)$ is analogous to Eq.~(\ref{eq:pinunu}),
except~\cite{BurasNLO} the change of $\kappa_+$ to $\kappa_L
=(2.12 \pm 0.03) \times 10^{-10}\times(|V_{us}|/0.224)^8$, and
taking only the imaginary part for the various CKM products. Since
$\phi_{ds} \sim -60^\circ$ or $+35^\circ$ have large imaginary
part, while $r_{ds} \equiv \vert V_{t^\prime d}^*V_{t^\prime
s}\vert \sim 5\times 10^{-4}$ is stronger than the SM3 expectation
of ${\rm Im}\, V_{td}^*V_{ts} \sim 10^{-4}$, we expect the CPV
decay rate of $K_L \to \pi^0\nu\bar\nu$ to be much enhanced.

We plot ${\cal B}(K_L \to \pi^0\nu\bar\nu)$ vs $\phi_{ds}$ in
Fig.~3(d), for $r_{ds} = 4$ and $6 \times 10^{-4}$. Reading off
from the figure, we see that the {\it $K_L \to \pi^0\nu\bar\nu$
rate can reach above $10^{-9}$}, almost two orders of magnitude
above SM3 expectation of $0.3\times 10^{-10}$. It is likely above
$5\times 10^{-10}$, and in general larger than $K^+\to
\pi^+\nu\bar\nu$.
Specifically, for our nominal value of $r_{ds} \sim 5\times
10^{-4}$ and $\phi_{ds} \sim +35^\circ$, ${\cal B}(K_L \to
\pi^0\nu\bar\nu)$ and ${\cal B}(K^+\to \pi^+\nu\bar\nu)$ are $6.5$
and $2 \times 10^{-10}$, respectively, while for the $\phi_{ds}
\sim -60^\circ$ case, they are $12$ and $3 \times 10^{-10}$,
respectively. The latter case is closer to the Grossman-Nir
bound~\cite{GrossNir}, i.e. ${\cal B}(K_L \to
\pi^0\nu\bar\nu)/{\cal B}(K^+\to \pi^+\nu\bar\nu) \sim
\tau_{K_L}/\tau_{K^+} \sim 4.2$, because $V_{t^\prime d}
V_{t^\prime s}^*$ is more imaginary.
Thus, both $K^+\to \pi^+\nu\bar\nu$ and $K_L \to \pi^0\nu\bar\nu$
should be very interesting at the next round of experiments.
We note that the ongoing E391A experiment could~\cite{Blucher}
attain single event sensitivity with the Grossman-Nir bound based
on the current ${\cal B}(K^+\to \pi^+\nu\bar\nu)$ measurement.
However, for $r_{ds} \sim 3.5 \times 10^{-4}$ and $\phi_{ds} \sim
-45^\circ$, which is still a solution for $R_6 \sim 2.2$, one has
${\cal B}(K_L \to \pi^0\nu\bar\nu) \sim 4\times 10^{-10}$ with
${\cal B}(K^+\to \pi^+\nu\bar\nu)$ at lower end of current range.

%\section{Check on $b\to d$? % Safe! (Non-trivial)
% }

With $\phi_{sb} \sim 70^\circ$ and $\phi_{ds} \sim -60^\circ$ (and
$+35^\circ$) both sizable while the associated CKM product is
larger than the corresponding SM3 top contribution, there is large
impact on $b\to s$ and $s\to d$ transitions from $Z$ penguin and
box diagrams. It is therefore imperative to check that one does
not run into difficulty with $b\to d$ transitions. Remarkably, we
find that the impact on $b\to d$ is mild.
From Eqs.~(\ref{eq:rsbphisb}), (\ref{eq:rdsphids}), (\ref{eq:s34})
and (\ref{eq:ds_vals}), we infer
\begin{equation}
r_{db} \sim 1\times 10^{-3},\ \ \phi_{db} \sim 10^\circ\
(105^\circ).
\label{eq:db_vals}
\end{equation}
Since $r_{db}$ is much smaller than $\vert V_{td}^*V_{tb}\vert
\sim \lambda^3 \sim 0.01$ in SM3, the impact on $b \to d$ is
expected to be milder, i.e. we are not far from the
$V_{t^{\prime}d} \to 0$ limit.
%This is particularly true for the relatively mild $\phi_{db} \sim 10^\circ$ case.
%
We stress that this is {\it nontrivial} since there is a large
effect in $b\to s$; it is a consequence of imposing $s\to d$ and
$Z\to b\bar b$ constraints.
We illustrate in Fig.~1(b) the unitarity quadrangle
\begin{equation}
V_{ud}V_{ub}^* + V_{cd}V_{cb}^* + V_{td}V_{tb}^* + V_{t^\prime
d}V_{t^\prime b}^* = 0.
\label{eq:dbquad}
\end{equation}
In contrast to Fig.~1(a), % for $b\to s$,
$(V_{td}V_{tb}^* + V_{t^\prime d}V_{t^\prime b}^*)_{\rm SM4}$ and
$(V_{td}V_{tb}^*)_{\rm SM3}$ can hardly be distinguished.

The $B^0_d$-$\overline B^0_d$ mass difference and CP violation
phase in mixing are respectively given by $\Delta m_{B_d}\equiv
2\left| M_{12} \right|$ and $\sin 2\Phi_{B_d}\equiv {\rm Im}\,
(M_{12}/|M_{12}|)$, where
\begin{eqnarray}
% B-B mixing
%\langle B\left|H_{\rm eff}^{\Delta B=2}\right|\overline B\rangle
M_{12} &=& \kappa_{B_d}
%\frac{G_F^2}{12 \pi^2}m_W^2 m_{B_q} B_{B_q} f_{B_q}^2 \eta_{\rm QCD}
\biggl[(\lambda_t^{db})^2 \eta_{t}S(x_t) +
(\lambda_{t^\prime}^{db})^2 \eta_{t^\prime}S(x_{t^\prime})
\nonumber
\\
& &
%+ (\lambda_{t^\prime}^{bs})^2 S(x_t^\prime)
+2 \lambda_t^{db} \lambda_{t^\prime}^{db}
\eta_{tt^\prime}S(x_t,x_{t^\prime}) \biggr],
\end{eqnarray}
with $\kappa_{B_d}=\frac{G_F^2}{12 \pi^2}m_W^2 m_{B_d} B_{B_d}
f_{B_d}^2$. The functions $S(x)$ and $S(x,y)$ can be found
in~\cite{BuchBurLaut}. We take $\eta_{t}=0.55$,
$\eta_{t^\prime}=0.58$ and $\eta_{tt^\prime}=0.50$, and plot in
Fig.~4(a) $\Delta m_{B_d}$ vs $\phi_{db}$, for $r_{db} = 8$ and
$12\times 10^{-4}$ (corresponding to $r_{ds} = 4$ and $6\times
10^{-4}$). We have taken the experimental value of $\Delta m_{B_d}
= (0.505\pm 0.005)$ ps$^{-1}$ from PDG 2005~\cite{PDG}, and
illustrated with the lower range of $f_{B_d} \sqrt{B_{B_d}} =
(246\pm 38)$ MeV~\cite{IStewart}. We have scaled up the error for
the latter by 1.4, since it comes from the new result on $f_{B_d}$
with unquenched lattice QCD~\cite{fB}, but $B_{B_d}$ is not yet
updated. We see from Fig.~4(a) that $\Delta m_{B_d}$ does not rule
out the parameter space around Eq.~(\ref{eq:db_vals}) (equivalent
to Eq.~(\ref{eq:ds_vals})). The overall dependence on $r_{db}$ and
$\phi_{db}$ is mild, and error on $f_{B_d} \sqrt{B_{B_d}}$
dominates. Seemingly, a lower value of $f_{B_d} \sqrt{B_{B_d}}
\sim 215$ MeV is preferred.
SM3 would give $\Delta m_{B_d} = 0.44 - 0.62$ ps$^{-1}$ for
$f_{B_d} \sqrt{B_{B_d}} = 208$ MeV $- 246$ MeV, so the problem is
not with SM4.

\begin{figure}[t!]
%\smallskip
%\vskip3.mm
\includegraphics[width=1.6in,height=1.08in,angle=0]{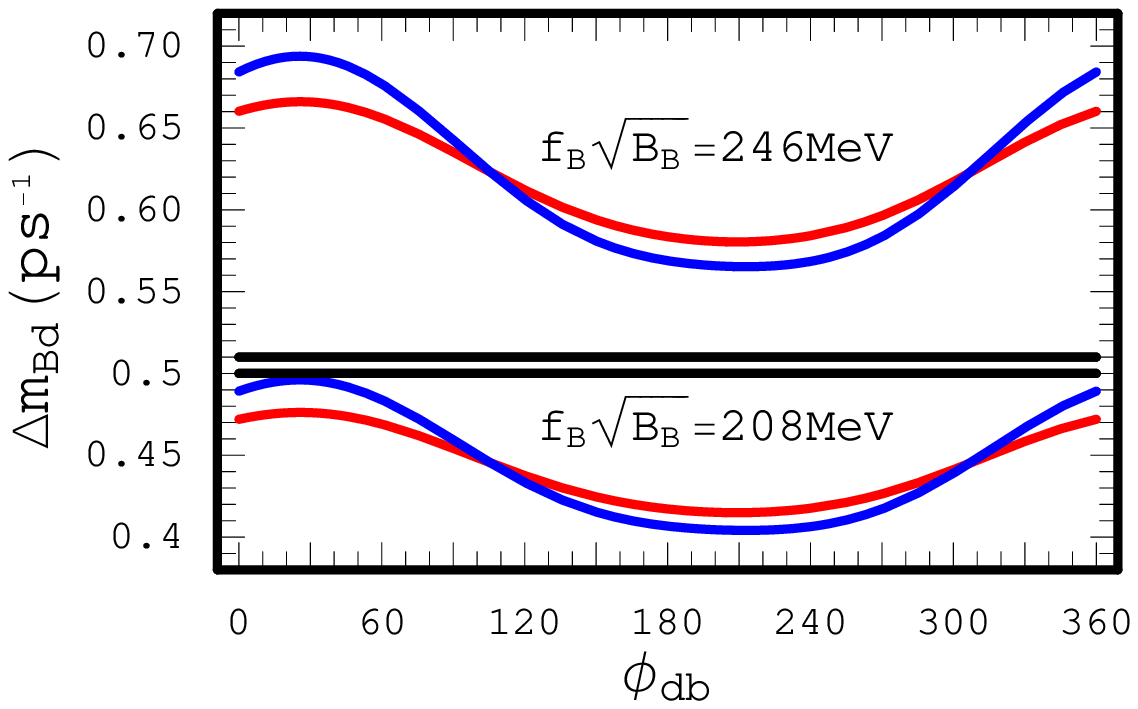}
%\vskip0.0mm
\includegraphics[width=1.6in,height=1.09in,angle=0]{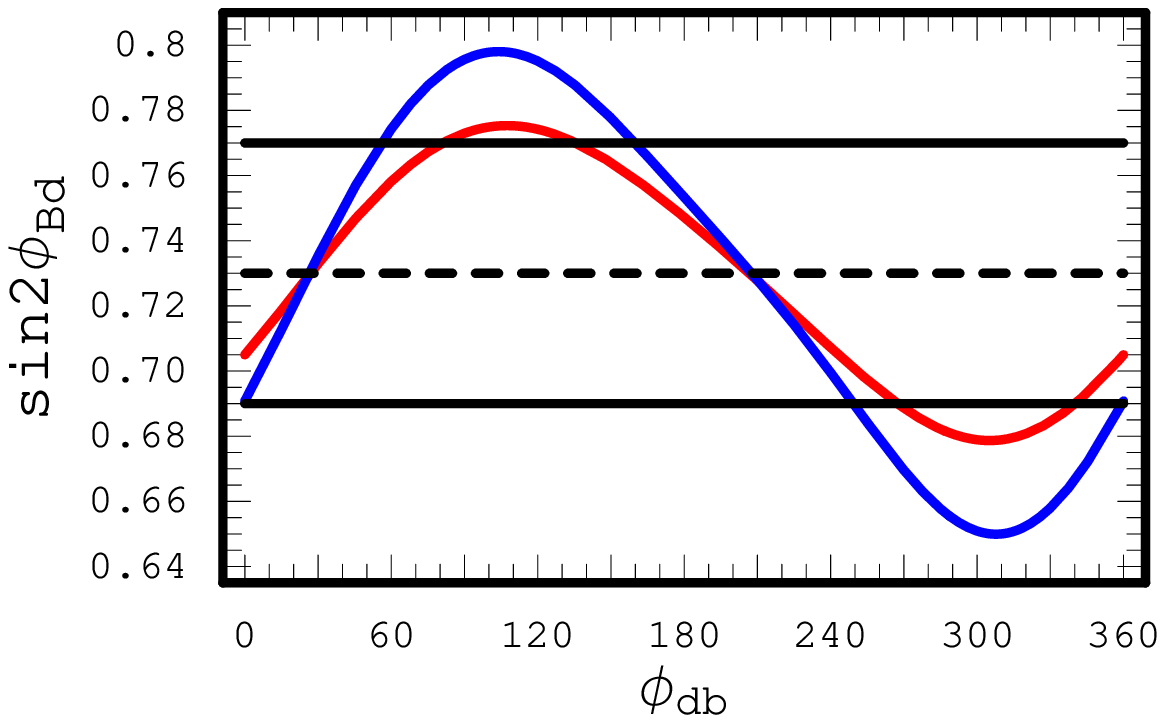}
%\smallskip\smallskip%\smallskip\smallskip\smallskip
\vskip1.mm
 \caption{
 (a) $\Delta m_{B_d}$ and (b) $\sin2\Phi_{B_d}$
   vs $\phi_{db}$ for $r_{db} = 8$ and $12\times 10^{-4}$, with
   $V^*_{t^\prime d}V_{t^\prime b} \equiv r_{db}\, e^{i\phi_{db}}$.
  Larger $r_{db}$ gives stronger variation, and horizontal bands are
  the experimental range~\cite{PDG}.
%  In (a), the two set of curves are for
%   $f_{B_d}\sqrt{B_{B_d}}$ values as marked.
 }
 \vskip3.mm
 \label{fig:Fig3}
\end{figure}

We plot $\sin2\Phi_{B_d}$ vs $\phi_{db}$ in Fig.~4(b), for $r_{db}
= 8$ and $12\times 10^{-4}$. One can see that $\sin2\Phi_{B_d}$,
which is not sensitive to hadronic parameters such as $f_{B_d}
\sqrt{B_{B_d}}$, is well within experimental range of
``$\sin2\phi_1$"$=0.73\pm0.04$ from PDG 2005~\cite{PDG} for the
$\phi_{db} \sim 10^\circ$ case.
However, for $\phi_{db} \sim 105^\circ$ case, which is much more
imaginary, $\sin2\Phi_{B_d}$ is on the high side~\cite{BelleLP05},
and it seems that CPV in $B$ physics prefers $R_6 \sim 2.2$ over
$R_6 \sim 1$.
As another check, we find the semileptonic asymmetry
%purely from $B^0$-$\bar B^0$ mixing,
$A_{SL} = -0.7\times 10^{-3}$ ($-0.2\times 10^{-3}$) for
$\phi_{db} \sim 10^\circ$ ($105^\circ$), which is also well within
range of $A_{SL}^{\rm exp} = (-1.1\pm7.9\pm7.0)\times
10^{-3}$~\cite{ASLBelle}.

%\section{$4\times 4$ Impression: Double Cabibbo? Unnatural?}
%

With Eqs.~(\ref{eq:rsbphisb}), (\ref{eq:s34}) and
(\ref{eq:db_vals}), together with standard (SM3) values for
$V_{cb}$ and $V_{ub}$, we can get a glimpse of the typical
$4\times 4$ CKM matrix, which appears like
\begin{equation}
\begin{scriptsize}
\left(
\begin{array}{cccc}
  0.9745  & 0.2225  & 0.0038\, e^{- i\, 60^\circ} &
  0.0281\, e^{ i\, 61^\circ}
 \\ \\
  - 0.2241  & 0.9667  & 0.0415  & 0.1164\, e^{ i\, 66^\circ}
 \\ \\
  0.0073\, e^{- i\, 25^\circ} & - 0.0555\, e^{-i\, 25^\circ} &
  0.9746  & 0.2168\, e^{- i\, 1^\circ}
  \\ \\
  - 0.0044\, e^{- i\, 10^\circ} & - 0.1136\, e^{- i\, 70^\circ} &
  - 0.2200  & 0.9688
\end{array}
\right)
\end{scriptsize}
 \label{eq:VCKM4}
\end{equation}
for $\phi_{db} \sim 10^\circ$ case ($V_{cd}$ and $V_{cs}$ pick up
tiny imaginary parts, which are too small to show in angles).
%
%\begin{equation}
%\begin{scriptsize}
%\left(
%\begin{array}{cccc}
%  0.9745  & 0.2225  & 0.0038\, e^{- i\, 60^\circ} &
%  0.0292\, e^{ i\, 74^\circ}
% \\ \\
%  - 0.2242  & 0.9667  & 0.0415  & 0.1161\, e^{ i\, 65^\circ}
% \\ \\
%  0.0082\, e^{- i\, 17^\circ} & - 0.0555\, e^{-i\, 25^\circ} &
%  0.9746  & 0.2168\, e^{- i\, 1^\circ}
%  \\ \\
%  0.0044\, e^{i\, 75^\circ} & - 0.1136\, e^{- i\, 70^\circ} &
%  - 0.2200  & 0.9688
%\end{array}
%\right),
%\end{scriptsize}
% \label{eq:VCKM4-105}
%\end{equation}
%
%The elements $V_{us}$, $V_{cb}$ and $V_{ub}$ (including phase of
%$-60^\circ$ are taken as input.
For the $\phi_{db} \sim 105^\circ$ case, the appearance is almost
the same, except $V_{td} \simeq 0.0082\, e^{-i\, 17^\circ}$ and
$V_{ub^\prime} \simeq 0.029\, e^{i\, 74^\circ}$.
% is not too different in strength).
Note the ``double Cabibbo" nature, i.e. the 12 and 34
diagonal $2\times 2$ submatrices appear almost the same.
% (except a larger imaginary part for $V_{tb^\prime}$).
This is a consequence of our choice of Eq.~(\ref{eq:s34}). To keep
Eq.~(\ref{eq:rsbphisb}) intact, however, weakening $s_{34}$ would
result in even large $V_{t^\prime s}$, but it would still be close
to imaginary.
Since $V_{t^{(\prime)}d}^*V_{t^{(\prime)}s}$ are tiny compared to
$V_{ud}^*V_{us} \simeq -V_{cd}^*V_{cs}$, the unitarity quadrangle
for $s\to d$ cannot be plotted as in Fig.~1. However, note that
$V_{td}^*V_{ts}$ is almost real, and CPV in $s\to d$ comes mostly
from $t^\prime$.
%, and would make the kaon constraint harder to fit ($R_8$ too small).
%Of course, the consequence of the strength of $V_{t^\prime s}$,
%and the resulting large impact on $V_{ts}$, are the salient
%features explored in this paper for $s\to d$ and $b\to d$
%transitions, leading to the $V_{t^\prime d}$ entry.

The entries for $V_{ib^\prime}$, $i = u$, $c$, $t$ are all
sizable. $\vert V_{ub^\prime}\vert \sim 0.03$ satisfies the
unitarity constraint $\vert V_{ub^\prime}\vert < 0.08$~\cite{PDG}
from the first row, but it is almost as large as $V_{cb}$.
However, the long standing puzzle of unitarity of the first row
could be taken as a hint for finite $\vert V_{ub^\prime}\vert \sim
0.03$~\cite{Vubp}.

The element $V_{cb^\prime} \simeq -V_{t^\prime s}^*$ is even
larger than $V_{cb}$ and close to imaginary. Together with finite
$V_{ub^\prime}$,  $V_{ub^\prime}V^*_{cb^\prime} \simeq 0.0033\,
e^{-i\, 5^\circ}$ ($0.0034\, e^{i\, 9^\circ}$) is not negligible,
and one may worry about $D^0$-$\bar D^0$ mixing. Fortunately the
$D$ decay rate is fully Cabibbo allowed.
%, while the CPV phase of $-5^\circ$ is much smaller
%than $\arg V_{t^\prime s}^* V_{t^\prime b}$.
%
Using $f_D\sqrt{B_D} = 200$ MeV, we find $\Delta m_{D^0} \lesssim
0.05$ ps$^{-1}$ for $m_{b^\prime} \lesssim 280$ GeV, for both
nominal cases of Eq.~(\ref{eq:db_vals}). Thus, the current bound
of $\Delta m_{D^0} < 0.07$ ps$^{-1}$ is satisfied, and the search
for $D^0$ mixing is of great interest. This bound weakens by
factor of 2 if one allows for strong phase between $D^0\to
K^-\pi^+$ and $K^+\pi^-$~\cite{PDG}.

If $m_{b^\prime} < m_{t^\prime}$, as slightly preferred by
$D^0$-$\bar D^0$ mixing constraint, the direct search for
$b^\prime$ just above 200 GeV at the Tevatron Run II could be
rather interesting. Since $V_{cb^\prime}$ is not suppressed, the
$b^\prime$ quark would decay via charged current. Both $b^\prime$
and $t^\prime$, regardless of which one is lighter, with
$m_{t^\prime} \sim 300$ GeV and $\vert m_{t^\prime} - m_{b^\prime}
\vert \lesssim 85$ GeV~\cite{PDG}, can be easily discovered at the
LHC.

%\section{Discussion and Conclusion}

The large and mainly imaginary element $V_{t^\prime s} \simeq
-V^*_{cb^\prime}$ in Eq.~(\ref{eq:VCKM4}), being larger than
$V_{ts}$ and $V_{cb}$, may appear unnatural (likewise for
$V_{ub^\prime}$ vs $V_{ub}$). However, it is allowed, since the
main frontier that we are just starting to explore is in fact
$b\to s$ transitions. The current situation that ${\cal
A}_{K^+\pi^-} \sim -0.12$ while ${\cal A}_{K^+\pi^0} \gtrsim 0$ in
$B \to K\pi$ decays may actually be hinting at the need for such
large $b\to s$ CPV effects.
The litmus test would be finding $\Delta m_{B_s}$ not far above
current bound, {\it but with sizable $\sin2\Phi_{B_s} <
0$}~\cite{Kpi0HNS}, which may even emerge at Tevatron Run II. Our
results studied here are for illustration purpose, but the main
result, that $K_L \to \pi^0\nu\bar\nu$ may be rather enhanced, is
a generic consequence of Eq.~(\ref{eq:rsbphisb}), which is a
possible solution to the $B^+\to K^+\pi^0$ DCPV puzzle.

In summary,
the deviation of direct CPV measurements between neutral and
charged $B$ decays, ${\cal A}_{K^+\pi^0} - {\cal A}_{K^+\pi^-}
\simeq 0.16$ while ${\cal A}_{K^+\pi^-} \simeq -0.12$, is a puzzle
that could be hinting at New Physics. A plausible solution is the
existence of a 4th generation with $m_{t^\prime} \sim 300$ GeV and
$V^*_{t^\prime s}V_{t^\prime b} \sim 0.025\, e^{i\,70^\circ}$. If
so, we find special solution space is carved out by stringent kaon
constraints, and the $4\times 4$ CKM matrix is almost fully
determined.
%, where, surprisingly, the hadronic parameters
%$R_6\gtrsim 2$, $R_8 \lesssim 1$ are preferred to be able to
%account for $\varepsilon^\prime/\varepsilon$.
$K^+\to \pi^+\nu\bar\nu$ may well be of order $(1-2)\times
10^{-10}$, while $K_L\to \pi^0\nu\bar\nu \sim (4-12)\times
10^{-10}$ is greatly enhanced by the large phase in $V^*_{t^\prime
d} V_{t^\prime s}$.
With kaon constraints satisfied, $B_d$ mixing and
$\sin2\Phi_{B_d}$ are consistent with experiment.
Our results are generic. If the effect weakens in $b\to s$
transitions, the effect on $K\to \pi\nu\bar\nu$ would also weaken.
But a large CPV effect in electroweak $b\to s$ penguins would
translate into an enhanced $K_L\to \pi^0\nu\bar\nu$ (and
$\sin2\Phi_{B_s} < 0$).

\vskip 0.3cm \noindent{\bf Acknowledgement}.\ This work is
supported in part by NSC-94-2112-M-002-035, NSC94-2811-M-002-053
and HPRN-CT-2002-00292. WSH thanks SLAC Theory Group for
hospitality.


\begin{thebibliography}{99}
% PRD: Phys. Rev. D {\bf nn}, ppppp (yyyy)
% PLB: Phys. Lett. B {\bf nn}, ppp (yyyy)
% PRL: Phys. Rev. Lett. {\bf nn}, ppp (yyyy)
% ZPC: Z. Phys. C {\bf nn}, ppp (yyyy)
% JHEP: J. High Energy Phys. 01, ppp (yyyy)
% EPJ: Eur. Phys. J. C {\bf nn}, ppp (yyyy)
% NPB: Nucl. Phys. {\bf Bnnn}, ppp (yyyy)
% {\it et al.}
%
\bibitem{AKpiAKpi0}
B.~Aubert {\it et al.}  [BaBar Collab.], Phys.\ Rev.\ Lett.\ {\bf
93}, 131801 (2004); Y.~Chao {\it et al.}  [Belle Collab.], {\it
ibid.}\ {\bf 93}, 191802 (2004); K.~Abe {\it et al.} [Belle
Collab.], hep-ex/0507045.
%
\bibitem{LargeC}
C.W. Chiang, M. Gronau, J.L. Rosner and D.A. Suprun, Phys. Rev. D
{\bf 70}, 034020 (2004); Y.Y. Charng and H.n.~Li,  {\it ibid.} D
{\bf 71}, 014036 (2005); M.~Gronau and J.L.~Rosner,  {\it ibid.}\
D {\bf 71}, 074019 (2005); C.S. Kim, S.~Oh and C. Yu,
hep-ph/0505060; H.n. Li, S. Mishima and A.I. Sanda,
hep-ph/0508041.
%
\bibitem{Kpi0HNS}
W.S. Hou, M. Nagashima and A. Soddu, hep-ph/0503072, to appear in
Phys.\ Rev.\ Lett.
%
\bibitem{PQCDKpiLO}
Y.Y. Keum, H.n. Li and A.I. Sanda, Phys. Rev. D {\bf 63}, 054008
(2001).
%
\bibitem{HSS87}
W.S.~Hou, A.~Soni and H.~Steger, Phys. Lett. B {\bf 192}, 441
(1987).
%
\bibitem{varphi3}
Our results do not change drastically as $\phi_{ub}$ is varied by
$\pm 10^\circ$.
%
\bibitem{DKDalitz}
A.~Giri {\it et al.}, % Y.~Grossman, A.~Soffer and J.~Zupan,
Phys.\ Rev.\ D {\bf 68}, 054018 (2003);
%K.~Abe {\it et al.}  [Belle Collab.], hep-ex/0308043;
A.~Poluektov {\it et al.}  [Belle Collab.], {\it ibid.}\ D {\bf
70}, 072003 (2004).
%
\bibitem{PDG}
S. Eidelman {\it et al.} [Particle Data Group], Phys. Lett. B {\bf
592}, 1 (2004); and 2005 update at http://pdg.lbl.gov/.
%
\bibitem{Yanir}
T.~Yanir, JHEP06 (2002) 044.
%
\bibitem{E787}
S.C.~Adler {\it et al.}  [E787 Collab.],
  Phys.\ Rev.\ Lett.\  {\bf 79}, 2204 (1997).

%
\bibitem{E949}
V.V. Anisimovsky {\it et al.} [E949 Collab.],
  Phys. Rev. Lett. {\bf 93}, 031801 (2004).
%
\bibitem{BSU}
  A.J.~Buras, F.~Schwab and S.~Uhlig,
  hep-ph/0405132.
%
\bibitem{E871}
D.~Ambrose {\it et al.}  [E871 Collab.],
  Phys.\ Rev.\ Lett.\  {\bf 84}, 1389 (2000).
%
\bibitem{eprime}
A.~Alavi-Harati {\it et al.}  [KTeV Collab.],
  Phys.\ Rev.\ Lett.\  {\bf 83}, 22 (1999);
V.~Fanti {\it et al.}  [NA48 Collab.],
  Phys.\ Lett.\ B {\bf 465}, 335 (1999).
%
\bibitem{BurasNLO}
A.J.~Buras and M.~Jamin, JHEP01 (2004) 048.
%
\bibitem{Bijnens}
J. Bijnens and J. Prades, JHEP06 (2000) 035.
%
%\bibitem{HPdR}
%  T.~Hambye, S.~Peris and E.~de Rafael,
%  JHEP {\bf 0305}, 027 (2003).
%  %[arXiv:hep-ph/0305104].
%
\bibitem{GrossNir}
Y.~Grossman and Y.~Nir,
  Phys.\ Lett.\ B {\bf 398}, 163 (1997).
%
\bibitem{Blucher}
E. Blucher, plenary talk at the XXII Lepton-Photon Symposium, June
2005, Uppsala, Sweden.
%
\bibitem{BuchBurLaut}
G.~Buchalla, A.J.~Buras and M.E.~Lautenbacher, Rev. Mod. Phys.
{\bf 68}, 1125 (1996).
%
\bibitem{IStewart}
I. Stewart, plenary talk at  the XXII Lepton-Photon Symposium,
June 2005, Uppsala, Sweden.
%
\bibitem{fB}
A.~Gray {\it et al.}  [HPQCD Collab.], hep-lat/0507015.
%
\bibitem{BelleLP05}
The summer 2005 result by K.~Abe {\it et al.} [Belle Collab.],
hep-ex/0507037, reports a low value of $\sin2\phi_1 = 0.652 \pm
0.039 \pm 0.020$, but this is for $B^0\to J/\psi K^0$ mode only,
and it is too early to draw any conclusion.
%
\bibitem{ASLBelle}
E. Nakano {\it et al.} [Belle Collab.], hep-ex/0505017. This new
result is in agreement with PDG 2005 with slightly better errors.
%
\bibitem{Vubp}
Recent kaon decay results imply a more stringent
bound~\cite{Blucher} of $1-\vert V_{ud}\vert^2-\vert V_{us}\vert^2
= 0.0004\pm0.0011$ ($\vert V_{ub}\vert^2$ is negligible), or
$\vert V_{ub^\prime}\vert < 0.047$ at 90\% C.L., which is still
satisfied by our value.
%
\end{thebibliography}
\end{document}